\documentclass[a4,12pt]{article}

\usepackage{amsmath,amssymb}

\def\tfontsize{scaled\magstep4}
\font\titlerm=cmr10 \tfontsize

\makeatletter

\@addtoreset{equation}{section}

\renewcommand{\section}{\@startsection{section}{1}{\z@}
	{-3.5ex \@plus -1ex \@minus -.2ex}{2.3ex \@plus.2ex}
	{\normalfont\normalsize\bfseries}}
\renewcommand{\subsection}{\@startsection{subsection}{2}{\z@}
	{-3.5ex \@plus -1ex \@minus -.2ex}{2.3ex \@plus.2ex}
	{\normalfont\normalsize\bfseries}}
\renewcommand{\subsubsection}{\@startsection{subsubsection}{3}{\z@}
	{-3.5ex \@plus -1ex \@minus -.2ex}{2.3ex \@plus.2ex}
	{\normalfont\normalsize\it}}
\makeatother

\newcommand\tr{\mathop{\rm tr}\nolimits}
\renewcommand\Im{\mathop{\rm Im}\nolimits}
\renewcommand\Re{\mathop{\rm Re}\nolimits}

\newcommand\rrangle{{\rangle\!\rangle}}
\newcommand\llangle{{\langle\!\langle}}

\newcommand\vk{{\vec k}}
\newcommand\vl{{\vec l\,}}
\newcommand\vp{{\vec p\,}}
\newcommand\vq{{\vec q\,}}

\newcommand\CN{{\cal N}}
\newcommand\CO{{\cal O}}
\newcommand\CP{{\cal P}}
\newcommand\CS{{\cal S}}
\newcommand\CT{{\cal T}}

\begin{document}
\begin{titlepage}
\begin{flushright}
{\tt arXiv:1602.00720}
\end{flushright}
\setlength{\baselineskip}{19pt}
\bigskip\bigskip\bigskip

\vbox{\centerline{\titlerm Entanglement Entropy of Scattering Particles}
}

\bigskip\bigskip\bigskip

\centerline{Robi Peschanski\footnote{\tt robi.peschanski@cea.fr}}
\medskip
{\it 
\centerline{Institut de Physique Th{\' e}orique, CEA-Saclay, F-91191 Gif-sur-Yvette, France}
}
\bigskip
\centerline{and}
\bigskip
\centerline{Shigenori Seki\footnote{\tt sigenori@hanyang.ac.kr}}
\medskip
{\it 
\centerline{Research Institute for Natural Sciences, Hanyang University}
\centerline{Seoul 133-791, Republic of Korea}
\centerline{Osaka City University Advanced Mathematical Institute (OCAMI)}
\centerline{3-3-138, Sugimoto, Sumiyoshi-ku, Osaka 558-8585, Japan}
}

\vskip .3in

\centerline{\bf abstract}

We study the entanglement entropy between the two outgoing particles 
in an elastic scattering process. 
It is formulated within an S-matrix formalism using  
the partial wave expansion of two-body states, which plays a 
significant role in our computation. 
As a result, we obtain a novel formula that expresses the entanglement 
entropy in a high energy scattering by the use of physical observables, 
namely the elastic and total cross sections and a physical bound on 
the impact parameter range, related to the elastic differential 
cross-section.

\vfill\noindent
1 February 2016
\end{titlepage}

\setlength{\baselineskip}{19pt}

\section{Introduction}

Entanglement is a significant concept which appears 
in various subjects of quantum physics. 
The quantum entanglement has been attracting many attentions 
of theoretical physicists, 
since remarkable progresses on the entanglement between the systems 
on two regions were made 
in quantum field theories \cite{CC} and holography \cite{RT} 
and the intriguing conjecture called ER=EPR \cite{MS} was suggested. 
In the context of the ER=EPR conjecture, 
the entanglements between two particles, 
which are, for example, a pair of accelerating quark and anti-quark \cite{JK} 
and a pair of scattering gluons \cite{SS}, have been studied. 
Then it naturally induces the following primitive question: 
How does the entanglement entropy of a pair of particles change 
from an initial state to a final one in an elastic channel of scattering process? 
It is qualitatively expected that the elastic collision of two initial particles, 
{\it e.g.}, in a high energy collider, generates some amount of entanglement 
between the particles in the final state. 
We are interested in quantifying the entanglement entropy generated by collision. 

By just neglecting inelastic channels in weak coupling perturbation \cite{BMV}, 
Ref.~\cite{SPS} analyzed such entanglement entropy in a field theory 
by the use of an S-matrix.\footnote{
We quote for completion Ref.~\cite{LBH}, where the entanglement entropy 
is discussed in a low energy decay process using different concept and method.} 
In this article we exploit the S-matrix formalism further 
in order for a non-perturbative understanding of the entanglement entropy 
in a scattering process with taking also an inelastic channel into account. 
This is especially required in the case of strong interaction scattering at high energy 
where inelastic multi-particle scattering contributes 
to a large part of the total cross-section, 
while elastic scattering is still important.  
The basic S-matrix formalism of strong interaction, 
as developed long time ago, {\it e.g.}, in Ref.~\cite{Va,BV}, 
allows us to find an approach to scattering processes
without referring explicitly to an underlying quantum field theory.  

Following Refs.~\cite{Va,BV}, we consider a scattering process 
of two incident particles, A and B, 
whose masses are $m_A$ and $m_B$ respectively, in 1+3 dimensions. 
This process is divided \cite{Va} into the following two channels: 
\begin{center}
\begin{tabular}{rl}
``Elastic'' channel: & A + B $\to$ A + B \\
``Inelastic'' channel: & A + B $\to$ X 
\end{tabular}
\end{center}
where X stands for any possible states except for the two-particle state, A + B. 
We postpone the study extended to a matrix including more varieties 
of two-particle channels \cite{BV} to a further publication.

The full Hilbert space of states is not usually factorized as 
${\cal H}_{\rm full} = {\cal H}_A \otimes {\cal H}_B \otimes {\cal H}_X$ in an interacting system. 
However the Hilbert space of both the initial and final states 
is factorizable in the S-matrix formalism, 
because one considers only asymptotic initial 
and final states long before and after the interaction. 
We introduce the S-matrix, ${\cal S},$ for the overall set of initial and final states. 
Once we fix an initial state $|{\rm ini}\rangle$, 
the final state $|{\rm fin}\rangle$ is determined by the S-matrix. 
In this article we are interested in the entanglement 
between two outgoing particles, A + B, 
in a final state of elastic scattering 
in the presence of a non-negligible fraction of open inelastic final states. 
Therefore we additionally introduce a projection operator $Q$ 
onto the two-particle Hilbert space ${\cal H}_A \otimes {\cal H}_B$ from ${\cal H}_{\rm full}$. 
Then the final elastic state, in other words, 
the state of two outgoing particles, is described as 
$|{\rm fin}\rangle = Q{\cal S} |{\rm ini}\rangle$. 

We employ the two-particle Fock space 
$\{|\vp\rangle\!{}_A\} \otimes \{|\vq\rangle\!{}_B\}$ 
as the Hilbert space ${\cal H}_A \otimes {\cal H}_B$. 
The two-particle state 
which consists of particle A with momentum $\vp$ and B with $\vq$ 
is denoted by $|\vp,\vq \rangle = |\vp\rangle\!{}_A \otimes |\vq\rangle\!{}_B$.
We define an inner product of the two-particle states in a conventional manner by 
$\langle \vp,\vq | \vk,\vl \rangle = 2E_{A\vp}\delta^{(3)}(\vp - \vk)\, 2E_{B\vq}\delta^{(3)}(\vq - \vl)$, 
where $E_{I\vp} = \sqrt{p^2 +m_I^2}$ $(I=A,B)$ and $p = |\vp|$.

We shall study the entanglement between the two outgoing particles, A and B. 
When the density matrix of the final state on ${\cal H}_A \otimes {\cal H}_B$ 
is denoted by $\rho$, 
we define a reduced density matrix as $\rho_A = \tr_B \rho$. 
Then the entanglement entropy is given by $S_{\rm EE} = -\tr_A \rho_A \ln \rho_A$. 
The other way to calculate the entanglement entropy is 
to use the R{\' e}nyi entropy, 
$S_{\rm RE}(n) = (1-n)^{-1}\ln \tr_A (\rho_A)^n$. 
It leads to the entanglement entropy described as 
$S_{\rm EE} = \lim_{n\to 1} S_{\rm RE}(n) = -\lim_{n\to 1}{\partial \over \partial n}\tr_A (\rho_A)^n$.

\section{Partial wave expansion}

The partial wave expansion is often useful to analyze a scattering process. 
Before starting to study the entanglement entropy, 
let us recall what Refs.~\cite{Va,BV} studied. 

We adopt a center-of-mass frame. 
The state of the two particles, $\mathrm{A} + \mathrm{B}$, 
which have momenta $\vp$ and $-\vp$ is denoted by 
$|\vp \rrangle := |\vp, -\vp\rangle$, 
while the many-particle state of X is denoted by $|X\rangle$. 
Since the complete set of states is given by the orthogonal basis, 
$\{|\vp\rrangle, |X\rangle \}$, 
one can describe the identity matrix as 
\begin{align}
{\bf 1} = \int {d^3\vp \over 2E_{A\vp} 2E_{B\vp}\delta^{(3)}(0)}|
\vp\rrangle \llangle \vp| 
	+ \int dX\, |X\rangle \langle X| \,. \label{eq:totHilbIDop}
\end{align}
We notice that $\delta^{(3)}(0)$ comes from 
$\llangle \vk|\vl \rrangle = 2E_{A\vk}\, 2E_{B\vk}\, \delta^{(3)}(\vk -\vl) \delta^{(3)}(0)$, 
due to our definition of the inner product of states.

One can expand the S-matrix elements in term of partial waves. 
Let us consider the S-matrix and T-matrix defined by $\CS = {\bf 1} +2i\CT$. 
The unitarity condition is $\CS^\dagger \CS = {\bf 1}$, 
which is equivalent to $i(\CT^\dagger -\CT) = 2 \CT^\dagger \CT$. 
Extracting the factor of energy-momentum conservation, 
we describe the T-matrix elements as 
\begin{align}
\llangle \vp|\CT|\vq\rrangle = \delta^{(4)}(P_\vp-P_\vq)\llangle \vp|{\bf t}|\vq \rrangle \,, \quad
\llangle \vp|\CT|X\rangle = \delta^{(4)}(P_\vp-P_X)\llangle \vp|{\bf t}|X \rangle \,. 
\end{align}
$P_\vp$ and $P_X$ are the total energy-momenta 
of $|\vp\rrangle$ and $|X\rangle$ respectively, 
which say $P_\vp = (E_{A\vp}+E_{B\vp},0,0,0)$. 

One introduces the overlap matrix $F_{\vp\vk}(k,\cos\theta)$, 
\begin{align}
F_{\vp\vk} = {2\pi k \over E_{A\vk} +E_{B\vk}}\int dX \llangle \vp |{\bf t}^\dagger |X\rangle \delta^{(4)}(P_X -P_\vk) \langle X|{\bf t} |\vk \rrangle  \,, \label{eq:OLfunc}
\end{align}
where $k$ and $\theta$ are defined by 
$\vp\cdot \vk = pk \cos\theta$ and  $k=p$. 
This matrix implies the contribution of the inelastic channel 
at the middle of the scattering process. 
The T-matrix element in the elastic channel and the overlap matrix are decomposed 
in terms of partial waves, 
\begin{align}
{\pi k \over E_{A\vk}+E_{B\vk}}\llangle \vp|{\bf t}|\vk \rrangle &= \sum_{\ell = 0}^\infty (2\ell +1) \tau_\ell(k) P_\ell(\cos \theta) \,, \label{eq:tPW} \\
F_{\vp\vk}(k,\cos\theta)&= \sum_{\ell = 0}^\infty (2\ell +1) f_\ell(k) P_\ell(\cos \theta) \,, \label{eq:OLfuncPW}
\end{align}
where $P_\ell(\cos \theta)$ are the Legendre polynomials.
Then one can rewrite the unitarity condition as 
\begin{align}
\Im \tau_\ell = |\tau_\ell|^2 + {f_\ell \over 2} \,. \label{eq:unitarityPW}
\end{align}
Using $s_\ell := 1 +2i\tau_\ell$, which comes from the partial wave expansion 
of the S-matrix element, 
\begin{align}
{\pi k  \over E_{A\vk}+E_{B\vk}}\llangle \vp| {\bf s}| \vk \rrangle 
= \sum_{\ell = 0}^\infty (2\ell +1) s_\ell P_\ell(\cos \theta)\,, \label{eq:sPW}
\end{align}
the unitarity condition is equivalent to $s_\ell^* s_\ell = 1 - 2f_\ell$. 
If there is not an inelastic channel, {\it i.e.} $f_\ell = 0$, 
then the unitarity condition is reduced to $s_\ell^* s_\ell =1$. 
A comment in order \cite{Va,BV} is that we can define a  
pseudo-unitary two-body S-matrix with partial wave components, 
$\omega_\ell^* \omega_\ell = 1$, by rescaling $s_\ell$ 
as $\omega_\ell := s_\ell / \sqrt{1-2f_\ell}\,$.

The partial wave expansion allows us to depict the integrated elastic cross section, 
the integrated inelastic cross section and the total cross section as 
\begin{align}
\sigma_{\rm el} &= {4\pi \over k^2} \sum_{\ell=0}^\infty (2\ell +1)|\tau_\ell|^2 \,, \quad 
\sigma_{\rm inel} = {2\pi \over k^2} \sum_{\ell=0}^\infty (2\ell +1) f_\ell \,, \quad
\sigma_{\rm tot}  = {4\pi \over k^2} \sum_{\ell=0}^\infty (2\ell +1) \Im \tau_\ell \,. \label{eq:crosssec}
\end{align}
The differential elastic cross section is 
\begin{align}
{d\sigma_{\rm el} \over dt} = {\pi \over k^4}\sum_{\ell,\ell'}(2\ell + 1)(2\ell' + 1)\tau_\ell \tau_{\ell'}^* P_\ell(\cos\theta)P_{\ell'}(\cos\theta)
= {|A|^2 \over 64\pi sk^2} \,, \label{eq:difcrosssec}
\end{align}
where $A(s,t)$ is the scattering amplitude, $s$ and $t$ the Mandelstam variables, 
and the scattering angle $\cos\theta = 1 + t / (2k^2)$.

\section{Entanglement entropy of two particles}

We consider two unentangled particles, A and B, 
with momenta $\vk$ and $\vl$ as incident particles. 
That is to say, we choose a single state as an initial state;
\begin{align} 
|{\rm ini}\rangle = |\vk,\vl \rangle = |\vk\rangle\!{}_A \otimes |\vl\rangle\!{}_B \,. \label{eq:inistate}
\end{align}
Here we have not taken the center-of-mass frame yet. 
Of course the entanglement entropy of the initial state vanishes. 
In terms of the S-matrix, the final state of two particles, 
$|{\rm fin}\rangle = Q \CS|{\rm ini}\rangle$, is described as 
\begin{align}
|{\rm fin}\rangle 
= \biggl(\int {d^3\vp \over 2E_{A\vp}}{d^3\vq \over 2E_{B\vq}}|\vp,\vq\rangle \langle \vp,\vq|\biggr) 
	\CS |\vk,\vl\rangle \,. 
\end{align}
Then we can define the total density matrix of the final state by 
$\rho := \CN^{-1}|{\rm fin}\rangle \langle {\rm fin}|$.
The normalisation factor $\CN$ will be determined later 
so that $\rho$ satisfies $\tr_A\tr_B \rho = 1$. 
Tracing out $\rho$ with respect to the Hilbert space of particle B, 
we obtain the reduced density matrix, $\rho_A := \tr_B\rho$, namely, 
\begin{align}
\rho_A = {1 \over \CN} &\int {d^3\vp \over 2E_{A\vp}} {d^3\vq \over 2E_{B\vq}} {d^3\vp' \over 2E_{A\vp'}} \bigl(\langle \vp,\vq|\CS|\vk,\vl\rangle \langle \vk,\vl|\CS^\dagger|\vp',\vq\rangle \bigr) |\vp\rangle\!{}_A {}_A\!\langle\vp'| \,. 
\end{align}

Now let us adopt the center-of-mass frame, which leads to $\vk+\vl = 0$. 
Then the initial state is $|{\rm ini}\rangle = |\vk\rrangle$, 
and the reduced density matrix becomes 
\begin{align}
\rho_A &= {1 \over \CN} \int {d^3\vp \over 2E_{A\vp}} 
	{\delta(0)\delta(p-k) \over 4k (E_{A\vk} +E_{B\vk})}
	\bigl|\llangle \vp |{\bf s}|\vk \rrangle \bigr|^2
	|\vp\rangle\!{}_A {}_A\!\langle\vp| \,, \label{eq:reddiagmat}
\end{align}
where ${\bf s} = {\bf 1} +2i{\bf t},$ and $\delta(0)$ stems from the 
modulus equality of the initial and final particles' momenta. 
By substituting \eqref{eq:sPW} into \eqref{eq:reddiagmat}, 
the normalization condition, $\tr_A \rho_A = 1$, 
fixes $\CN$ as $\CN = \delta^{(4)}(0)\, \CN'$ with 
\begin{align}
\CN' = {E_{A\vk} + E_{B\vk} \over \pi k} \sum_{\ell=0}^\infty (2\ell +1)|s_\ell|^2 \,. 
\end{align}
Since $\tr_A (\rho_A)^n$ straightforwardly provides us the R{\' e}nyi and entanglement entropy, we calculate
\begin{align}
\tr_A (\rho_A)^n &= \int_{-1}^1 d\zeta\,  \CP(\zeta) G^{n-1}(\zeta) \,, \quad (\zeta := \cos\theta) \label{eq:trArhon} \\ 
\CP(\zeta) &= {1 \over 2}
	{\bigl| \sum_\ell (2\ell+1)s_\ell P_\ell(\zeta) \bigr|^2 \over \sum_\ell (2\ell+1)|s_\ell|^2} \,, \quad
G(\zeta) = 
	{\bigl| \sum_\ell (2\ell+1)s_\ell P_\ell(\zeta) \bigr|^2 \over \sum_\ell (2\ell+1)\cdot \sum_\ell (2\ell+1)|s_\ell|^2} \,, \label{eq:Probs}
\end{align}
where we used the three-dimensional Dirac delta function 
in spherical coordinates with azimuthal symmetry, 
$\delta^{(3)}(\vp-\vk) = (4\pi k^2)^{-1} \delta(p-k)\sum_\ell (2\ell +1) P_\ell(\cos\theta)$, 
and the partial wave expansion of a delta function, 
$2\delta(1-\cos\theta) = \sum_\ell (2\ell+1) P_\ell(\cos\theta)$. 
Due to $s_\ell = 1 + 2i\tau_\ell$ and the unitarity condition \eqref{eq:unitarityPW}, 
one can rewrite ${\cal P}(\zeta)$ in Eqs.~\eqref{eq:Probs} as 
\begin{align}
\CP(\zeta) = \delta(1-\zeta){V -4\sum_\ell(2\ell+1)\Im \tau_\ell \over V - 2\sum_\ell(2\ell+1)f_\ell} 
+{2|\sum_\ell(2\ell +1)\tau_\ell P_\ell(\zeta)|^2 \over V - 2\sum_\ell(2\ell+1)f_\ell} \,, \label{eq:Pzeta}
\end{align}
where $V := \sum_\ell (2\ell +1)$. 
Here  $\sum_\ell (2\ell +1) \Im \tau_\ell$, $\sum_\ell (2\ell +1) f_\ell$ 
and $|\sum_\ell (2\ell +1) \tau_\ell P_\ell(\zeta)|^2$ 
correspond to physical observables and thus are necessarily finite, 
while the infinite sum $V$ diverges. 
Therefore Eq.~\eqref{eq:Pzeta} leads to $\CP(\zeta) = \delta(1-\zeta)$. 
Then one can easily proceed the integration in \eqref{eq:trArhon} and gets finally
\begin{align}
\tr_A (\rho_A)^n &= K^{n-1} \,, \label{eq:trArhonKnminusone}\\ 
K &:= G(1) = {\bigl| \sum_\ell (2\ell+1)s_\ell \bigr|^2 \over \sum_\ell (2\ell +1) \cdot \sum_\ell (2\ell+1)|s_\ell|^2} \,. \label{eq:Ksell}
\end{align}
Obviously Eq.~\eqref{eq:trArhonKnminusone} for $n = 1$ correctly 
reproduces the normalization condition, $\tr_A\rho_A = 1$.

From Eq.~\eqref{eq:trArhonKnminusone} the R{\' e}nyi entropy 
is $S_{\rm RE} = -\ln K$ and equal to the entanglement entropy, 
\begin{align}
S_{\rm EE} = - \lim_{n \to 1} {\partial \over \partial n} \tr_A (\rho_A)^n = -\ln K \,. \label{eq:lnKernel}
\end{align}
Using a Cauchy-Schwarz inequality applied to \eqref{eq:Ksell}, 
$K$ satisfies $0 \leq K \leq 1$, 
that is to say, the entanglement entropy $S_{\rm EE}$ is equal to zero or positive.

When there is no interaction, $s_\ell$ is equal to one for all $\ell$ 
and $K$ becomes one, that is to say, 
the entanglement entropy $S_{\rm EE}$ vanishes. 
This is natural, because the final state is same as the initial state 
without interaction and the initial state \eqref{eq:inistate} is not entangled. 
On the other hand, if the system has interaction, 
the entanglement entropy is expected to increase in scattering processes.

We have a comment on the elastic case without the inelastic channel, 
{\it i.e.}, $f_\ell =0$ for all $\ell$. 
In this case, one has $s_\ell = \exp(2i\delta_\ell)$, 
where $\delta_\ell$ are the phase shifts. 
Hence one obtains the expression of Eq.~\eqref{eq:Ksell} in terms of the phase shifts, 
$K = V^{-2}(|\sum_\ell(2\ell+1)\cos 2\delta_\ell|^2 +|\sum_\ell(2\ell+1)\sin 2\delta_\ell|^2)$.

Let us rewrite Eq.~\eqref{eq:Ksell} in terms of $\tau_\ell$ and $f_\ell$ as 
\begin{align}
K = 1 - {4 \sum_\ell(2\ell+1)|\tau_\ell|^2 -{4 \over V} |\sum_\ell(2\ell+1)\tau_\ell|^2 \over V -2 \sum_\ell(2\ell+1)f_\ell} \,. \label{eq:Ktauf}
\end{align}
Formally the full Hilbert space extends over all partial waves, 
and thus one has $V = \sum_{\ell=0}^\infty (2\ell+1) = \infty$. 
It causes $K=1$, in other words, the entanglement entropy vanishes. 
However, in physical elastic processes, the Hilbert space is essentially limited 
by energy-momentum conservation, 
so that the physical Hilbert space provides a meaningful entanglement entropy 
as we shall see further.

Since the partial wave expansions of the integrated elastic cross section, 
the integrated inelastic cross section, the total cross section  
and the differential cross section are shown 
in Eqs.~\eqref{eq:crosssec} and \eqref{eq:difcrosssec}, 
$K$ can be described in terms of these physical observables as 
\begin{align}
K = 1- { \sigma_{\rm el} -{4k^2 \over V}{d\sigma_{\rm el} \over dt}\bigr|_{t=0} \over {\pi V \over k^2} -\sigma_{\rm inel} } \,. \label{eq:Kinfinit}
\end{align}
By a power expansion of $S_{\rm EE}$ with respect to $V^{-1} \ll 1$, 
we obtain $S_{\rm EE} = (k^2/\pi)\sigma_{\rm el} V^{-1} +\CO(V^{-2})$. 
The leading term is proportional to the elastic cross section, 
and this is consistent with the result in Ref.~\cite{SPS}, 
which calculated the entanglement entropy of two outgoing particles 
in the field theories in weak coupling perturbation.

\section{Physical Hilbert space}

In an actual scattering process at a given momentum $k$, 
too high angular momentum modes are strongly depleted 
and negligible in the elastic scattering amplitude. 
In a semi-classical picture using the impact parameter $b = \ell/k$ representation, 
the limitation can be depicted as a maximal sizable value $b/2 \le R$, 
where $R$ is interpreted as the mean of incident particle effective radii.
In this context the largest relevant angular momentum $\ell_{\rm max}$ is 
\begin{align}
\ell_{\rm max} \sim  2k R \,. \label{eq:ellmax}
\end{align}
In practice, we shall consider \eqref{eq:ellmax} 
as the maximal value of the angular momentum beyond 
which the summation over partial wave amplitudes $\tau_\ell $ can be neglected. 
We thus approximate by truncation the sum over $\ell$ of the Hilbert space states. 
Note that reasonable values of $R$ may be obtained 
from experimental determination of the impact-parameter profile of the scattering amplitude, 
which can be inferred \cite{MSM} from the elastic differential cross-section 
$d\sigma_{\rm el}/dt$.

At high energy, {\it i.e.}, large momentum $k$ with maximal impact parameter $2R$, 
$\ell_{\rm max}$ is large. 
Although the key point in the derivation of Eqs.~\eqref{eq:trArhonKnminusone} and \eqref{eq:Ksell} is that ${\cal P}(\zeta)$ is identified 
with the delta function coming from $\sum_{\ell=0}^\infty (2\ell+1)P_\ell(\zeta)$, 
it keeps approximately valid for large $\ell_{\rm max}$. 
Therefore under this approximation 
one can conclude the entropy is $S_{\rm RE} = S_{\rm EE} = -\ln K$ 
in replacing $\sum_{\ell=0}^\infty$ with $\sum_{\ell=0}^{\ell_{\rm max}}$. 
The Hilbert space volume becomes 
$V = \sum_{\ell=0}^{\ell_{\rm max}}(2\ell+1) = (1 + \ell_{\rm max})^2 \sim \ell_{\rm max}^2 \gg 1$. 
Then  Eq.~\eqref{eq:Kinfinit} remains a good approximation 
with the parameter $V /k^2 \sim 4R^2$. 
Finally $K$ is obtained as
\begin{align}
K \sim 1
-{\sigma_{\rm el} -{1 \over R^2}{d\sigma_{\rm el} \over dt}\bigr|_{t=0} \over 4\pi R^2 -\sigma_{\rm inel}} \,, \label{eq:KHighfinit}
\end{align}
so that one gets a finite value for the R{\' e}nyi and entanglement entropy. 
In this expression the explicit $V$ dependence disappears. 
Note that $4\pi R^2$ can be considered as the  classical ``geometric'' cross section of the scattering.
Formula \eqref{eq:KHighfinit} implies that, if we measure the cross sections 
and get an evaluation of the impact parameter profile in a collider experiment, 
one can give a reliable approximate estimate of the entanglement entropy 
of the final elastic state of the two outgoing particles.

It is instructive to examine the limiting values 
of \eqref{eq:KHighfinit} in $0 \leq K \leq 1$. 
The value $K = 1$, corresponding to zero entanglement entropy, 
can be met when $R^2$ reaches its minimal value $(d\sigma_{\rm el} / dt)|_{t=0}/\sigma_{\rm el}$, 
which is nothing else than the average size of the elastic diffraction peak. 
The limit $K\to 0$ ({\it i.e.}, $S_{\rm EE}\to \infty$) 
may be reached only at a zero of the expression 
$4\pi R^2-\sigma_{\rm tot}+ {{(d\sigma_{\rm el} / dt)|_{t=0}}/ R^2}$, 
whose only solution is  $\Re A(s, 0) =0$ and $\sigma_{\rm tot}=8\pi R^2$, 
that is twice the geometric cross section. 
An exception is when both numerator and denominator in \eqref{eq:KHighfinit} 
tend simultaneously to zero, namely 
$\sigma_{\rm el}=\sigma_{\rm inel}=\sigma_{\rm tot}/2 = 4\pi (d\sigma_{\rm el}/dt)|_{t=0}/\sigma_{\rm el} = 4\pi R^2$. 
Interestingly enough it corresponds to the so-called ``black disk'' limit, 
which happens to be phenomenologically relevant for the high energy asymptotics \cite{BDHH}.

\section{Conclusion and comments}

We have studied the entanglement entropy between two outgoing 
particles, A and B, in an elastic scattering at high energy, 
where many inelastic channels are also opened. 
In the derivation of the entanglement entropy, 
we used the unitarity condition on the S-matrix \eqref{eq:unitarityPW}.
As a result, we obtained the formula for the entanglement entropy \eqref{eq:lnKernel}, 
$S_{\rm EE} = -\ln K$, with Eq.~\eqref{eq:Ksell}.

The R{\' e}nyi entropy is same as the entanglement entropy, 
{\it i.e.}, $S_{\rm RE} = -\ln K$. 
This implies that the outgoing particles are maximally entangled. 
This is caused by the fact that 
the reduced density matrix \eqref{eq:reddiagmat} is diagonal 
due to the momentum conservation of two scattering particles 
in the center-of-mass frame.

Eq.~\eqref{eq:lnKernel} is reminiscent of Boltzmann's entropy formula 
with the Boltzmann constant $k_B = 1$. 
In this sense, one can regards $1/K$ as a kind of micro-canonical ensemble 
of final states. 
Indeed it can be recast in the following form derived from \eqref{eq:trArhon}:
\begin{align}
S_{\rm EE} 
= \ln {V \over 2} -\int_{-1}^1 d\zeta\  {\cal P}(\zeta) \ln {\cal P}(\zeta) \,,
\label{entropy}
\end{align}
due to ${\cal P(\zeta)} = {V \over 2} G(\zeta)$. 
${\cal P}(\zeta)$ is positive and of norm one 
in both cases of the full and physical Hilbert spaces, 
because $\int_{-1}^1 d\zeta\, {\cal P}(\zeta) = 1$ 
thanks to the orthogonality of Legendre polynomials. 
Hence one can identify ${\cal P}(\zeta)$ 
with a well-defined probability measure over the interval $\zeta \in [-1,+1]$. 
We also see that ${\cal P}(\zeta)$ originates from 
the probability $|\llangle \vp |{\bf s}|\vk \rrangle|^2$ in Eq.~\eqref{eq:reddiagmat}. 
Since $V$ can be interpreted as the total number 
of final two-body quantum states ($2\ell +1$ at level $\ell$), 
the second term in Eq.~\eqref{entropy} can be understood of the 
correction to the total entropy due to entanglement.

The result for $K$ is described as Eqs.~\eqref{eq:Ktauf} and \eqref{eq:Kinfinit}. 
The subspace volume of elastic states is small in size with respect to 
the volume of the overall Hilbert space,  $K$ is almost equal to one. 
In other words, the entanglement entropy is negligibly small.

For scattering at high energy, conveniently 
called ``soft scattering'', 
we can employ the physical truncation of the Hilbert space given by Eq.~\eqref{eq:ellmax}. 
We take the limit of large momentum $k$ 
with a fixed maximal impact parameter $2R$. 
Then $K$ becomes Eq.~\eqref{eq:KHighfinit}. 
This implies that the entanglement entropy is described 
in terms of the cross sections and the maximal impact parameter. 
Since it is possible to measure these parameters in experiments, 
{\it e.g.}, a proton-proton scattering in a collider, 
the entanglement entropy can be evaluated using \eqref{eq:KHighfinit}. 
It would be interesting, in order to confirm the validity of our formula, 
to confront this result obtained within the S-matrix framework of strong interactions, 
to a microscopic derivation of the entanglement entropy 
in a gauge field theory at strong coupling using, {\it e.g.}, 
the AdS/CFT correspondence. 
It would require the holographic study of a QCD-like theory.

\bigbreak\bigskip\bigskip
\centerline{{\bf Acknowledgments}}\nobreak

SS was supported in part by the National Research Foundation of Korea grant No.~NRF-2013R1A1A2059434 and No.~NRF-2013R1A2A2A05004846. 
SS is grateful to Institut des Hautes {\' E}tudes Scientifiques (IH{\' E}S) for their hospitality. 

\end{document}